\documentclass[reprint, aip, jcp, amsmath,amssymb]{revtex4-1}

\usepackage{graphicx}
\usepackage{dcolumn}
\usepackage{bm}
\usepackage{color}
\usepackage{enumitem}
\usepackage[normalem]{ulem}
\usepackage{comment}

\newif\ifHighlitedChanges
\def\ifHighlitedChanges{\iftrue}
\ifHighlitedChanges
  
  \def\STRIKE#1{{\color{red}\sout{#1}}}
\else
  
  \def\STRIKE#1{\relax}
\fi

\begin{document}
\title{Molecular heat transport across a time-periodic temperature gradient}
\author{Renai Chen}
\affiliation{Theoretical Division and Center for Nonlinear Studies, Los Alamos National Laboratory, Los Alamos, New Mexico}
\author{Tammie Gibson}
\affiliation{Theoretical Division, Los Alamos National Laboratory, Los Alamos, New Mexico}
\author{Galen T. Craven}
\email{galen.craven@gmail.com}
\affiliation{Theoretical Division, Los Alamos National Laboratory, Los Alamos, New Mexico}

\begin{abstract}
The time-periodic modulation of a temperature gradient can alter the heat transport properties of a physical system.
Oscillating thermal gradients give rise to 
behaviors 
such as 
modified thermal conductivity
and
controllable time-delayed energy storage that are not present in a system with static temperatures.
Here, we examine how the heat transport properties of a  molecular lattice model are affected by an oscillating temperature gradient.
We use analytical analysis and molecular dynamics simulations to investigate the vibrational heat flow in a molecular lattice system consisting of a chain of particles connected to two heat baths at different temperatures, where the temperature difference between baths is oscillating in time.
We derive expressions for heat currents in this system using a stochastic energetics framework and a nonequilibrium Green's function approach that is modified to treat the nonstationary average energy fluxes.
We find that emergent energy storage, energy release, and thermal conductance mechanisms induced by the temperature oscillations can be controlled by varying the frequency, waveform, and amplitude of the oscillating gradient. 
The developed theoretical approach provides a general framework to describe how vibrational heat transmission through a molecular lattice is affected by temperature gradient oscillations.
\end{abstract}

\maketitle

\section{Introduction}

Many natural and technological systems derive their function from energy transport processes.
\cite{Li2012, Segal2016,Sabhapandit2012,Lebowitz1959,Lebowitz1967,Lebowitz1971,Lebowitz2008,Lebowitz2012,Nitzan2003thermal,Segal2005prl,Lebowitz2012,Dhar2015,Velizhanin2015,Esposito2016,craven16c,matyushov16c,craven17a,craven17b,craven17e,craven18b,craven20a,craven21a,chen2020local,chen2023jcp,Ochoa2022}
At the nanoscale, energy transport often involves a complex interplay and competition between multiple mechanisms and physical processes. 
For example, in nanotechnologies, energy transport may involve vibrational (phononic), electronic, and radiative mechanisms.
The development of theoretical methods to understand nanoscale heat transport has a rich history. \cite{Lebowitz1959,Lebowitz1967,Lebowitz1971,Cahill2003,Segal2005prl,Segal2016,Nitzan2007,Sato2012,Maldovan2013,Leitner2008,Leitner2013,Seifert2015periodictemp,Li2012,Dubi2011,Lim2013,Volz2022, HernandezJPCL2023, Krivtsov2023} 
Most studies of nanoscale energy transport are conducted at steady state under conditions in which the system distributions and the driving thermodynamic forces (such as a temperature gradient) are static in time. 
This gives rise to average energy fluxes that are constant in time. 
There are, however, a number of interesting and emergent energy transport properties that occur when a system is driven from steady state by a time-dependent force such as a time-varying temperature gradient.
In these cases, the system will generally not reach a steady state, and will instead relax to a time-dependent nonequilibrium state. 
In this article, we investigate the vibrational heat flow in such a system, specifically in a model molecular lattice structure connecting two heat baths with different temperatures, where the temperature difference between baths is oscillating in time. 

At the nanoscale, heat transport properties often violate Fourier's law (FL)---the primary physical principle that is used to understand heat transport at the macroscale.\cite{BonettoFourier2000,Bonetto2004SCR,Segal2009SCR,Chang2008,craven2023a} 
The core issue that induces the deviations is the interplay between diffusive and ballistic transport mechanisms, the latter being a mechanism that is not typically prominent in macroscale heat transport. 
Ballistic transport will generally not satisfy FL behavior, which is derived from a diffusive linear response picture. 
Although it is important to note that it cannot be stated generally that if a system’s
transport mechanism is diffusive, FL behavior will be observed.
In fact, nanoscale systems that are dominated by diffusive transport often do not follow FL. \cite{craven2023a} 
The physical structure of a system and its environment must often be purposefully engineered
to generate FL behavior.\cite{BonettoFourier2000,Bonetto2004SCR,Segal2009SCR}
Nanoscale heat transport must therefore be treated using other theoretical principles.

The first experimental measurements of single molecule thermal conductance have recently been performed. \cite{Reddy2019nature,Mosso2019} 
These measurements have confirmed anomalous thermal transport behavior with respect to FL over molecular length scales and
have also opened opportunities to engineer and then realize novel transport properties at the molecular level.
A common experimental setup for probing single-molecule transport is a molecular junction, a system consisting of 
a molecular bridge connecting two electrodes. 
Transport through the molecular bridge, both electronic and thermal, can be induced by applying a voltage bias and/or temperature bias to the junction setup.
Molecular junctions have been studied extensively
and are well-used to probe molecular transport mechanisms. \cite{Reddy2007,Tan2011,Lee2013,Kim2014,Venkataraman2015,Garner2018,Reddy2019nature,Mosso2019,Zimbovskaya2020}

From a theoretical standpoint, lattice models
provide tractable and useful mathematical frameworks to examine molecular thermal conductance. \cite{Isaeva2019modeling,Lepri2003,BonettoFourier2000,Bonetto2004SCR,Segal2009SCR,Lebowitz1959,Lebowitz1967,Lebowitz1971,Segal2021,segal2003thermal,Simine2010FPUT,CampbellFPUT2005,Dhar2008,dhar2006heat}
Lattice models have been instrumental in the study of heat transport and energy distribution at the nanoscale 
where the thermal conductance of molecules and materials can differ significantly from macroscale behaviors as discussed above.
For example, the deviation from FL behavior at the nanoscale was predicted well before the 
recent experimental set-ups capable of measuring single molecule thermal transport were developed. \cite{BonettoFourier2000,Bonetto2004SCR,Segal2009SCR,Chang2008,Lepri2003}

Typical examinations of molecular heat transport are conducted under the condition of a static temperature gradient, i.e., a temperature gradient that does not change in time. 
Time-dependent temperature gradient oscillations can alter a system's heat transport properties in comparison to the static case.
Theoretical frameworks to understand systems with time dependent temperatures have been developed, \cite{Reimann2002,brey1990generalized,hern07a,hern13d,Ford2015,Seifert2015periodictemp,Seifert2016periodiccurrent,Awasthi2021,Portugal2022effective,Volz2022,Weron2022,Ben-Abdallah2017thermalmemristor,Ordonez-Miranda2019thermalmemristor,Krivtsov2020} and we 
have developed an approach to describe heat transport in such systems. \cite{craven2023b}
In practical settings, temperature oscillations can be achieved using various methods such as applying a thermal source
that switches on and off at periodic intervals, using a laser to heat the system in a periodic manner, applying flash heating techniques, and/or by modulating the system pressure. \cite{Gruebele2012,Gruebele2013,platkov2014periodic,Wang2007}

We have previously examined the heat conduction properties of a single particle in contact with two thermal baths, 
where the temperature difference between the baths is oscillating in time. \cite{craven2023b}
We found that complex energy flux hysteresis and energy storage processes were induced by the temperature oscillations.
These effects have potential applications in, for example,
thermoelectric and pyroelectric materials, \cite{Bowen2014pyro,Yamamoto2021,Lheritier2022pyro,Dubi2011}
thermal batteries, \cite{Gur2012thermalbattery,Wang2022battery}
and energy transport devices. \cite{bartussek1994periodically,Hanggi2009wireperiodictemps,Zhang2008ratchet}
Here, we extend our previous work on a single particle system by examining a molecular chain consisting of multiple interacting particles in the presence of an oscillating temperature gradient.

The remainder of the article is organized as follows: 
Section~\ref{sec:model} contains the details of the molecular model we use to examine energy transport across an oscillating temperature gradient.  
In Sec.~\ref{sec:formalism}, the definitions and general formalism for the energy fluxes and thermal conductance in the model are presented.
Section~\ref{sec:hc} contains derivations of the energy flux expressions for the system and the thermal baths.
Section~\ref{sec:results} contains results and discussion about how the energy storage, energy release, and thermal conductance mechanisms induced by the temperature oscillations can be controlled by varying properties of the oscillating gradient.
Conclusions and implications of the work are presented in Sec.~\ref{sec:conclusions}.

\section{Model Details and Formalism}

\begin{figure*}
	\includegraphics[width=\textwidth]{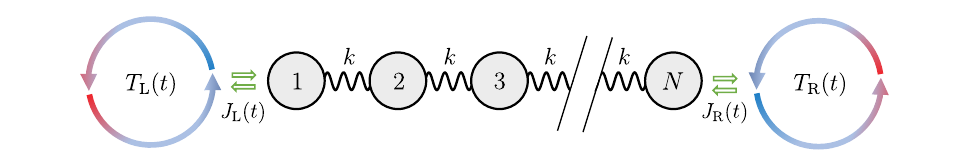}
	\caption{Schematic diagram of an example molecular chain with $N$ particles connecting two heat baths with oscillating temperatures. 
	The temperatures of the left and right baths, $T_\text{L}(t)$ and $T_\text{R}(t)$, are time-dependent and oscillate periodically with different frequencies.
	Because of the oscillating temperature gradient, 
	the energy fluxes $J_\text{L}(t)$ and $J_\text{R}(t)$ in/out of the two baths are also time-dependent and oscillatory. Adjacent particles in the chain are connected by a harmonic potential with force constant $k$.}
	\label{fig:schematic}
\end{figure*}

\label{sec:model}

The model we consider is a molecular chain, i.e., a one-dimensional lattice, consisting of $N$ particles with equal masses connected by harmonic interactions.
The first particle in the chain (the leftmost particle, i.e, particle $1$) is coupled to a heat bath $\text{L}$ with time-dependent oscillatory temperature $T_\text{L}(t)$ and the last particle (the rightmost particle, i.e, particle $N$) is coupled to a heat bath $\text{R}$ with time-dependent oscillatory temperature $T_\text{R}(t)$.
A schematic diagram of the model is shown in Fig.~\ref{fig:schematic}.
In order to distinguish the heat baths from the molecular chain, we will refer to the chain as the ``system'' and the heat baths as ``baths''.
In the limit of constant temperatures in the baths, $T_\text{L}(t) = T^{(0)}_\text{L}$ and $T_\text{R}(t) = T^{(0)}_\text{R}$, this model becomes 
a paradigmatic and well-studied model for heat transport in nanoscale systems.
Here, we take the temperatures to oscillate in time, and show that these oscillations give rise to transport phenomena that are not present in the limit of a static temperature gradient.

We will work in the limit that the system dynamics can be described classically. The Langevin equations of motion for this system are:
\begin{equation}
\begin{aligned}
\label{eq:EoM1}
m \ddot{x}_1 &= -(k + k_\text{pin}) x_1 + k x_2 - m \gamma_\text{L} \dot{x}_1 + \xi_\text{L}(t), \\
\ldots \\
m \ddot{x}_j  &= -2kx_j+kx_{j-1}+kx_{j+1}, \\
\ldots \\
m \ddot{x}_N &= -(k + k_\text{pin}) x_N + k x_{N-1} - m \gamma_\text{R} \dot{x}_N + \xi_\text{R}(t),
\end{aligned}
\end{equation}
where $m$ is the particle mass (taken to be the same for every particle), $x_j$ is the displacement of particle $j$ from its equilibrium position,
$k$ is the force constant between particles,
$k_\text{pin}$ is the force constant for a pinning potential for the two edge particles,
$\gamma_\text{L}$ and $\gamma_\text{L}$ are coupling constants between the system and the respective baths, and 
$\xi_\text{L}(t)$ and $\xi_\text{R}(t)$ are stochastic forces that obey the correlations
\begin{equation}
\begin{aligned}
\label{eq:FD_theorem_t}
 \big\langle \xi_\text{L}(t) \xi_\text{L}(t')\big\rangle &= 2 \gamma_\text{L}m k_\text{B} T_\text{L}(t)\delta(t-t'), \\
 \big\langle \xi_\text{R}(t) \xi_\text{R}(t')\big\rangle &= 2 \gamma_\text{R}m k_\text{B} T_\text{R}(t)\delta(t-t'), \\
	\big\langle \xi_\text{L}(t) \xi_\text{R}(t')\big\rangle &= 0, \\
 \big\langle \xi_\text{L}(t)\big\rangle &=0,\\
 \big\langle \xi_\text{R}(t)\big\rangle &=0,
\end{aligned}
\end{equation}
where $k_\text{B}$ is the Boltzmann constant and $T_\text{L}(t)$ and $T_\text{R}(t)$ are the time-dependent temperatures of the left and right bath, respectively.
We will consider the case where the temperatures of each bath take the specific forms:
\begin{equation}
\begin{aligned}
\label{eq:t-dep_T}
 T_\text{L}(t)   &= T^{(0)}_\text{L} + \Delta T_\text{L} \sin(\omega_\text{L} t),\\[1ex]
 T_\text{R}(t)   &= T^{(0)}_\text{R} + \Delta T_\text{R} \sin(\omega_\text{R} t),
\end{aligned}
\end{equation}
where $T^{(0)}_\text{L}$ and $T^{(0)}_\text{R}$ are the temperatures of the two baths in the limit of vanishing of oscillations, $\Delta T_\text{L}$ and $\Delta T_\text{R}$ define the amplitude of the oscillations, and $\omega_\text{L}$ and $\omega_\text{R}$ are oscillation frequencies. 
For ease of exposition, we define a temperature parameter for the system as 
\begin{equation}
T = \frac{\gamma_\text{L}T^{(0)}_\text{L} + \gamma_\text{R}T^{(0)}_\text{R}}{\gamma_\text{L} + \gamma_\text{R}},
\end{equation}
but it is important to note that because of the oscillating gradient there is no local temperature in the chain. 
The friction coefficients in the model used here are static, but memory effects can be included. \cite{Dhar2008,hern07a, hern13d}
We consider here only cases in which $\omega_\text{L}$ and $\omega_\text{R}$ are commensurate so that the system is periodic with total period $\mathcal{T}$.

\subsection{\label{sec:formalism} Energy Flux Formalism}
The oscillating temperature gradient in the system gives rise to time-dependent energy fluxes in the system
Using the stochastic energetics formalism of Sekimoto, \cite{Sekimoto1998}
the energy fluxes in our model can be separated into  three terms:
\begin{enumerate}
  \item $J_\text{sys}$ is the energy flux in/out of the system  
  \item $J_\text{L}$ is the the energy flux associated with the left bath 
  \item $J_\text{R}$ is the the energy flux associated with the right bath
\end{enumerate}
It is important to note that in the presence of static temperature gradient under steady-state conditions,
the $J_\text{sys}$ term goes to zero and the heat current through the system is defined by $J_\text{L} = - J_\text{R}$.
The principal goal of this manuscript is to describe how an oscillating temperature gradient affects these
energy fluxes in a harmonic lattice and to understand how those energy fluxes in turn alter thermal conductance.
The energy fluxes for the system, left bath, and right bath can be defined as, 
\begin{align}
\label{eq:general_heat}
J_\text{sys} (t) &= \partial_t \big\langle E(t)\big\rangle, \\[1ex]
\label{eq:heatcurrentbathR}
J_\text{L} (t) &=  m \gamma_\text{L} \big\langle \dot{x}_1^2(t)\big\rangle - \big\langle\xi_\text{L}(t) \dot{x}_1(t)\big\rangle,\\[1ex]
\label{eq:heatcurrentbathS}
J_\text{R} (t) &=  m \gamma_\text{R} \big\langle \dot{x}_N^2(t)\big\rangle - \big\langle \xi_\text{R}(t) \dot{x}_N(t)\big\rangle,
\end{align}
where $E(t)$ is the total energy of the system and $\langle  \rangle$ represents an ensemble average. 
A more general form of energy fluxes will be developed below.

Because of the temperature oscillations, the system does not reach a steady state.
Instead, the system will approach a periodic nonequilibrium state in the long-time limit ($t \to \infty)$.
This means that the system energy flux $J_\text{sys} (t)$ does not go to zero when the temperatures of the baths are time-dependent. Compare this with the case in which the temperature gradient is static and the system reaches a steady state in the long-time limit. In this case, $J_\text{sys} (t) = 0$.
The implication of this is that in the periodic nonequilibrium state, the system can store and then release energy due to the temperature oscillations.
These storage/release properties are prominent when the temperature oscillation frequencies are fast relative to the relaxation rates into the baths parameterized by $\gamma_\text{L}$ and $\gamma_\text{R}$. 
We define the average energy storage capacity of the system through the property, 
\begin{equation}
\label{eq:heatplus}
I^+_\text{sys} =  \frac{1}{\mathcal{T}^+}\int_0^{\mathcal{T}} J_\text{sys} (t') \Theta[J_\text{sys} (t') ]  dt'
\end{equation}
where $\Theta$ is the Heaviside function and
\begin{equation}
\label{eq:timeplus}
\mathcal{T}^+ = \int_0^{\mathcal{T}}  \Theta[J_\text{sys} (t') ]  dt'
\end{equation}
is the amount of time over an oscillation period the system heat current is positive.
The metric $I^+_\text{sys}$ measures 
the average positive system energy flux
and is directly related to
the amount of energy stored by the system over a period of oscillation.
The $I^+_\text{sys}$ and $\mathcal{T}^+$ integrals are taken over the time interval  $[0,\mathcal{T}]$, i.e., over a period of total system oscillation.

\section{\label{sec:hc} Molecular Heat Transport}

Our principal goal is to develop analytical expressions to understand heat transport through the harmonic chain in the prescence of an oscillating temperature gradient.
Harmonic interactions dominate the heat transport properties in a multitude of systems, and therefore we expect that the 
developed theory will accurately describe a number of molecular systems. 
However, the inclusion of anharmonic interactions between particles can be important and may give rise to new phenomena, but is beyond scope of this work.

To derive the energy fluxes, we will work in Fourier space. 
Transforming the equations of motion in Eq.~(\ref{eq:EoM1}), the system of equations in frequency space becomes
\begin{equation}
\begin{aligned}
\label{eq:Langevin_chain_FT}
-m\omega^2\tilde x_1 &= - (k+k_\text{pin})\tilde x_1 + k\tilde x_2  -i\omega m \gamma_\text{L}\tilde x_1 + \tilde \xi_\text{L}, \\
\ldots \\
-m\omega^2\tilde x_j &= -k(2\tilde x_j-\tilde x_{j-1}-\tilde x_{j+1}), \\
\ldots \\
-m\omega^2\tilde x_N &= - (k+k_\text{pin})\tilde x_N + k\tilde x_{N-1} -i\omega m \gamma_\text{R}\tilde x_N+\tilde\xi_\text{R}.
\end{aligned}
\end{equation}
where $\tilde x$ and $\tilde\xi$ are the respective Fourier-transformed representations of $x$ and $\xi$.

Nonequilibrium Green's Function (NEGF) approaches for phononic transport~\cite{Yamamoto2006nonequilibrium}
can be employed to evaluate the energy fluxes.\cite{dhar2006heat}
The application of the NEGF formalism to compute transport properties is a common and well-established approach to understand
heat conduction in molecular structures under steady-state conditions.\cite{Dhar2008,segal2003thermal,Segal2016,sharony2020stochastic,Klockner2017}
Here, because the examined system does not approach a steady state, in order to include time-dependent properties of the temperature oscillations
we will need to modify the standard NEGF formalism using a bottom-up approach from the equations of motion. Developing this modified NEGF method will allow us 
to derive expressions for the energy fluxes.

Using matrix notation, the equations of motion for the particles in the molecular chain can be written as
\begin{align}
\label{eq:Langevin_chain_FT_M}
\mathbf{\tilde X}(\omega)&=\boldsymbol{G}(\omega)\mathbf{\Xi}(\omega),\\
\boldsymbol{G}(\omega)&=\left[-\mathbf{M}\omega^2+\mathbf{\Phi} + i \boldsymbol{\Gamma}_\text{L}(\omega) + i \boldsymbol{\Gamma}_\text{R}(\omega)\right]^{-1},
\end{align}
where $\tilde{\mathbf{X}}^T=\{\tilde x_1, \tilde x_2, \cdots, \tilde x_N\}$ is a Fourier-transformed displacement vector obtained using the relation $\mathbf{X}(t)=\int_{-\infty}^{\infty}d\omega\tilde{\mathbf{X}}(\omega)e^{i\omega t}$.
The stochastic force vectors are $\mathbf{\Xi}_\text{L}^T=\{\tilde\xi_\text{L},0, 0, \cdots \}$ and $\mathbf{\Xi}_\text{R}^T=\{0,0,\cdots,\tilde\xi_\text{R}\}$ and we define
$\mathbf{\Xi}(\omega)=\mathbf{\Xi}_\text{L}(\omega)+\mathbf{\Xi}_\text{R}(\omega)$ for notational simplicity.
The function $\boldsymbol{G}(\omega)$ is a Green's function with conjugate adjoint $\boldsymbol{G}^\dag(\omega)$. 
In the Green's function, $\textbf{M}$ is a diagonal mass matrix, $\boldsymbol{\Phi}$ a force constant matrix extracted from Eq.~(\ref{eq:EoM1}), and 
$\boldsymbol{\Gamma}_\text{L}(\omega)$ and $\boldsymbol{\Gamma}_\text{R}(\omega)$ are spectral functions of the heat baths that connect to the molecular system.
For Ohmic baths,\cite{Dhar2008} as we will consider here,
the spectral functions are $N \times N$ matrices
with element $\boldsymbol{\Gamma}_\text{L}(\omega)_{00}=m\omega\gamma_\text{L}$ 
and element $\boldsymbol{\Gamma}_\text{R}(\omega)_{NN}=m\omega\gamma_\text{R}$,
while all other elements in the matrices are zero.

Applying the oscillating temperatures from Eq.~(\ref{eq:t-dep_T}), 
the noise-noise correlations in Fourier space are 
\begin{widetext}
\begin{align}
\label{eq:fluctuation_correlationL}
 \nonumber \big\langle \tilde\xi_\text{L}(\omega) \tilde\xi_\text{L}(\omega')\big\rangle &= 
 \frac{ \gamma_\text{L}m k_\text{B}}{2\pi^2}\int_{-\infty}^{\infty}T_\text{L}(t)e^{-i(\omega+\omega^\prime)t}dt \\ 
 &=\frac{ \gamma_\text{L}m k_\text{B}}{\pi}\left[T^{(0)}_\text{L}\delta(\omega+\omega')
 +i\frac{\Delta T_\text{L}}{2}\delta(\omega_\text{L}+\omega+\omega') 
 - i \frac{\Delta T_\text{L}}{2}\delta(\omega_\text{L}-\omega-\omega')\right], \\
 \label{eq:fluctuation_correlationR}
 \nonumber \big\langle \tilde\xi_\text{R}(\omega) \tilde\xi_\text{R}(\omega')\big\rangle 
 &= \frac{ \gamma_\text{R}m k_\text{B}}{2\pi^2}\int_{-\infty}^{\infty}T_\text{R}(t)e^{-i(\omega+\omega^\prime)t}dt \\
 &= \frac{ \gamma_\text{R}m k_\text{B}}{\pi}\left[T^{(0)}_\text{R}\delta(\omega+\omega')
 +i\frac{\Delta T_\text{R}}{2}\delta(\omega_\text{R}+\omega+\omega') 
 -i \frac{\Delta T_\text{R}}{2}\delta(\omega_\text{R}-\omega-\omega')\right], 
\end{align}
\end{widetext}
where we have used the Fourier definition of the Dirac delta function and the complex exponential form for the trigonometric function to express the final forms for the correlations.
The cross correlation terms are$\big\langle \tilde\xi_\text{L}(\omega) \tilde\xi_\text{R}(\omega')\big\rangle = \big\langle \tilde\xi_\text{R}(\omega) \tilde\xi_\text{L}(\omega')\big\rangle = 0$.
Defining $\mathbf{\Lambda}_\text{L}=\mathbf{\Gamma}_\text{L}/\omega$ and $\mathbf{\Lambda}_\text{R}=\mathbf{\Gamma}_\text{R}/\omega$, the correlation functions between the stochastic noise vectors are
\begin{align}
 &\nonumber \big\langle \mathbf{\Xi}_\text{L}(\omega) \mathbf{\Xi}_\text{L}^T(\omega')\big\rangle = \frac{k_\text{B}T^{(0)}_\text{L}\mathbf{\Lambda}_\text{L}}{\pi}\delta(\omega+\omega') \\
 \nonumber \qquad  &-\frac{k_\text{B}\Delta T_\text{L}\mathbf{\Lambda}_\text{L}}{2\pi i} \big[\delta(\omega_\text{L}+\omega+\omega') - \delta(\omega_\text{L}-\omega-\omega')\big],\\
& \nonumber \big\langle \mathbf{\Xi}_\text{R}(\omega) \mathbf{\Xi}_\text{R}^T(\omega')\big\rangle = \frac{k_\text{B}T^{(0)}_\text{R}\mathbf{\Lambda}_\text{R}}{\pi}\delta(\omega+\omega') \\
\nonumber \qquad &-\frac{k_\text{B}\Delta T_\text{R}\mathbf{\Lambda}_\text{R}}{2\pi i}\big[\delta(\omega_\text{R}+\omega+\omega') - \delta(\omega_\text{R}-\omega-\omega')\big]. \\
\end{align}

We will now evaluate the energy fluxes for the left and right baths by combining the noise-noise correlations, the NEGF formalism, and the stochastic energetics expressions for the general energy flux expressions.
Using the equations of motion in Eq.~(\ref{eq:Langevin_chain_FT_M}), and applying inverse Fourier transforms, we can write expressions for the correlation functions needed to evaluate the energy fluxes. 
In matrix notation, the left and right energy fluxes can be expressed as
\begin{align}
	\label{eq:current_left}
	J_\text{L}(t) &= \left\langle (\mathbf{\Lambda}_\text{L}\dot{\mathbf{X}})^T \dot{\mathbf{X}}\right\rangle - \left\langle \dot{\mathbf{X}}^T\mathbf{\Xi}_\text{L}\right\rangle, \\
 \label{eq:current_right}
 	J_\text{R}(t) &= \left\langle (\mathbf{\Lambda}_\text{R}\dot{\mathbf{X}})^T \dot{\mathbf{X}}\right\rangle - \left\langle \dot{\mathbf{X}}^T\mathbf{\Xi}_\text{R}\right\rangle.
\end{align}
Therefore, deriving expressions for the energy fluxes of the baths requires evaluation of the correlation functions in Eqs.~(\ref{eq:current_left}) and (\ref{eq:current_right}).

The first term on the RHS of Eq.~(\ref{eq:current_left}) and can be evaluated as
\begin{widetext}
\begin{align}
\label{eq:v_v_l}
	\nonumber \left\langle(\mathbf{\Lambda}_\text{L}\dot{\mathbf{X}})^T \dot{\mathbf{X}} \right\rangle
	&=-\int_{-\infty}^{\infty}\omega e^{i\omega t}d\omega\int_{-\infty}^{\infty}\omega'e^{i\omega' t} d\omega' \left\langle \text{Tr}[\mathbf{\Xi}^T(\omega')\mathbf{G}^T(\omega')\mathbf{\Lambda}_\text{L}^T\mathbf{G}(\omega)\mathbf{\Xi}(\omega)] \right\rangle, \\ \nonumber
	&=\int_{-\infty}^{\infty}\omega^2 \frac{k_\text{B}T^{(0)}_\text{L}}{\pi}d\omega\text{Tr}[\mathbf{G}^\dagger(\omega)\mathbf{\Lambda}_\text{L}^T\mathbf{G}(\omega)\mathbf{\Lambda}_\text{L}]  +\int_{-\infty}^{\infty}\omega^2 \frac{k_\text{B}T^{(0)}_\text{R}}{\pi}d\omega\text{Tr}[\mathbf{G}^\dagger(\omega)\mathbf{\Lambda}_\text{L}^T\mathbf{G}(\omega)\mathbf{\Lambda}_\text{R}]  \\ \nonumber
	&-\int_{-\infty}^{\infty}\omega (\omega+\omega_\text{L})e^{-i\omega_\text{L} t}\frac{k_\text{B}\Delta T_\text{L}}{2\pi i}d\omega\text{Tr}[\mathbf{G}^\dagger(\omega+\omega_\text{L})\mathbf{\Lambda}_\text{L}^T\mathbf{G}(\omega)\mathbf{\Lambda}_\text{L}]  \\ \nonumber
	&+\int_{-\infty}^{\infty}\omega (\omega-\omega_\text{L})e^{i\omega_\text{L} t}\frac{k_\text{B}\Delta T_\text{L}}{2\pi i}d\omega\text{Tr}[\mathbf{G}^\dagger(\omega-\omega_\text{L})\mathbf{\Lambda}_\text{L}^T\mathbf{G}(\omega)\mathbf{\Lambda}_\text{L}]  \\ \nonumber
	&-\int_{-\infty}^{\infty}\omega (\omega+\omega_\text{R})e^{-i\omega_\text{R} t}\frac{k_\text{B}\Delta T_\text{R}}{2\pi i}d\omega\text{Tr}[\mathbf{G}^\dagger(\omega+\omega_\text{R})\mathbf{\Lambda}_\text{L}^T\mathbf{G}(\omega)\mathbf{\Lambda}_\text{R}]  \\ 
	&+\int_{-\infty}^{\infty}\omega (\omega-\omega_\text{R})e^{i\omega_\text{R} t}\frac{k_\text{B}\Delta T_\text{R}}{2\pi i}d\omega\text{Tr}[\mathbf{G}^\dagger(\omega-\omega_\text{R})\mathbf{\Lambda}_\text{L}^T\mathbf{G}(\omega)\mathbf{\Lambda}_\text{R}],  
\end{align}
\end{widetext}
and correspondingly the first term on the RHS of Eq.~(\ref{eq:current_right}) can be evaluated in a similar way yielding 
\begin{widetext}
\begin{align}
\label{eq:v_v_r}
	\nonumber \left\langle(\mathbf{\Lambda}_\text{R}\dot{\mathbf{X}})^T \dot{\mathbf{X}} \right\rangle
	&=-\int_{-\infty}^{\infty}\omega e^{i\omega t}d\omega\int_{-\infty}^{\infty}\omega'e^{i\omega' t} d\omega' \left\langle \text{Tr}[\mathbf{\Xi}^T(\omega')\mathbf{G}^T(\omega')\mathbf{\Lambda}_\text{R}^T\mathbf{G}(\omega)\mathbf{\Xi}(\omega)] \right\rangle, \\ \nonumber
	&=\int_{-\infty}^{\infty}\omega^2 \frac{k_\text{B}T^{(0)}_\text{R}}{\pi}d\omega\text{Tr}[\mathbf{G}^\dagger(\omega)\mathbf{\Lambda}_\text{R}^T\mathbf{G}(\omega)\mathbf{\Lambda}_\text{R}]  +\int_{-\infty}^{\infty}\omega^2 \frac{k_\text{B}T^{(0)}_\text{L}}{\pi}d\omega\text{Tr}[\mathbf{G}^\dagger(\omega)\mathbf{\Lambda}_\text{R}^T\mathbf{G}(\omega)\mathbf{\Lambda}_\text{L}]  \\ \nonumber
	&-\int_{-\infty}^{\infty}\omega (\omega+\omega_\text{R})e^{-i\omega_\text{R} t}\frac{k_\text{B}\Delta T_\text{R}}{2\pi i}d\omega\text{Tr}[\mathbf{G}^\dagger(\omega+\omega_\text{R})\mathbf{\Lambda}_\text{R}^T\mathbf{G}(\omega)\mathbf{\Lambda}_\text{R}]  \\ \nonumber
	&+\int_{-\infty}^{\infty}\omega (\omega-\omega_\text{R})e^{i\omega_\text{R} t}\frac{k_\text{B}\Delta T_\text{R}}{2\pi i}d\omega\text{Tr}[\mathbf{G}^\dagger(\omega-\omega_\text{R})\mathbf{\Lambda}_\text{R}^T\mathbf{G}(\omega)\mathbf{\Lambda}_\text{R}]  \\ \nonumber
	&-\int_{-\infty}^{\infty}\omega (\omega+\omega_\text{L})e^{-i\omega_\text{L} t}\frac{k_\text{B}\Delta T_\text{L}}{2\pi i}d\omega\text{Tr}[\mathbf{G}^\dagger(\omega+\omega_\text{L})\mathbf{\Lambda}_\text{R}^T\mathbf{G}(\omega)\mathbf{\Lambda}_\text{L}]  \\ 
	&+\int_{-\infty}^{\infty}\omega (\omega-\omega_\text{L})e^{i\omega_\text{L} t}\frac{k_\text{B}\Delta T_\text{L}}{2\pi i}d\omega\text{Tr}[\mathbf{G}^\dagger(\omega-\omega_\text{L})\mathbf{\Lambda}_\text{R}^T\mathbf{G}(\omega)\mathbf{\Lambda}_\text{L}],  
\end{align}
\end{widetext}
where we have used the identity, $\mathbf{G}^T(-\omega) = \mathbf{G}^\dagger(\omega)$.
The second terms (the noise-velocity correlations) on the RHS of Eqs.~(\ref{eq:current_left}) and (\ref{eq:current_right}) can be evaluated as
\begin{widetext}
\begin{align}
\label{eq:v_xi_l}
	\nonumber \left\langle\dot{\mathbf{X}}^T \mathbf{\Xi}_\text{L} \right\rangle
	&=i\int_{-\infty}^{\infty}e^{i\omega t}d\omega\int_{-\infty}^{\infty}\omega' e^{i\omega' t} d\omega' \left\langle \text{Tr}[\mathbf{\Xi}^T(\omega')\mathbf{G}^T(\omega')\mathbf{\Xi}_\text{L}(\omega)] \right\rangle \\ 
	&=-i\frac{k_\text{B}T_\text{L}(t)}{\pi}\int_{-\infty}^{\infty}\omega d\omega\text{Tr}[\mathbf{\Lambda}_\text{L}\mathbf{G}^\dagger(\omega)], \\
\label{eq:v_xi_r}
	\nonumber \left\langle\dot{\mathbf{X}}^T \mathbf{\Xi}_\text{R} \right\rangle
	&=i\int_{-\infty}^{\infty}e^{i\omega t}d\omega\int_{-\infty}^{\infty}\omega' e^{i\omega' t} d\omega' \left\langle \text{Tr}[\mathbf{\Xi}^T(\omega')\mathbf{G}^T(\omega')\mathbf{\Xi}_\text{R}(\omega)] \right\rangle \\ 
	&=-i\frac{k_\text{B}T_\text{R}(t)}{\pi}\int_{-\infty}^{\infty}\omega d\omega\text{Tr}[\mathbf{\Lambda}_\text{R}\mathbf{G}^\dagger(\omega)]. 
\end{align}
\end{widetext}
Combining these correlation functions we now arrive at the exact expressions for the time-dependent energy fluxes of the left heat bath, the right heat bath, and the system:
\begin{widetext}
\begin{align}
\label{eq:left_flux}
	\nonumber J_\text{L}(t) 
	=&\int_{-\infty}^{\infty}\omega^2 \frac{k_\text{B}T^{(0)}_\text{L}}{\pi}d\omega\text{Tr}[\mathbf{G}^\dagger(\omega)\mathbf{\Lambda}_\text{L}^T\mathbf{G}(\omega)\mathbf{\Lambda}_\text{L}] +\int_{-\infty}^{\infty}\omega^2 \frac{k_\text{B}T^{(0)}_\text{R}}{\pi}d\omega\text{Tr}[\mathbf{G}^\dagger(\omega)\mathbf{\Lambda}_\text{L}^T\mathbf{G}(\omega)\mathbf{\Lambda}_\text{R}]  \\  \nonumber
	&-\int_{-\infty}^{\infty}\omega (\omega+\omega_\text{L})e^{-i\omega_\text{L} t}\frac{k_\text{B}\Delta T_\text{L}}{2\pi i}d\omega\text{Tr}[\mathbf{G}^\dagger(\omega+\omega_\text{L})\mathbf{\Lambda}_\text{L}^T\mathbf{G}(\omega)\mathbf{\Lambda}_\text{L}]  \\ \nonumber
 	&-\int_{-\infty}^{\infty}\omega (\omega+\omega_\text{R})e^{-i\omega_\text{R} t}\frac{k_\text{B}\Delta T_\text{R}}{2\pi i}d\omega\text{Tr}[\mathbf{G}^\dagger(\omega+\omega_\text{R})\mathbf{\Lambda}_\text{L}^T\mathbf{G}(\omega)\mathbf{\Lambda}_\text{R}]  \\ \nonumber
	&+\int_{-\infty}^{\infty}\omega (\omega-\omega_\text{L})e^{i\omega_\text{L} t}\frac{k_\text{B}\Delta T_\text{L}}{2\pi i}d\omega\text{Tr}[\mathbf{G}^\dagger(\omega-\omega_\text{L})\mathbf{\Lambda}_\text{L}^T\mathbf{G}(\omega)\mathbf{\Lambda}_\text{L}]  \\ \nonumber
	&+\int_{-\infty}^{\infty}\omega (\omega-\omega_\text{R})e^{i\omega_\text{R} t}\frac{k_\text{B}\Delta T_\text{R}}{2\pi i}d\omega\text{Tr}[\mathbf{G}^\dagger(\omega-\omega_\text{R})\mathbf{\Lambda}_\text{L}^T\mathbf{G}(\omega)\mathbf{\Lambda}_\text{R}]  \\ 
	&+i\frac{k_\text{B} T_\text{L}(t)}{\pi}\int_{-\infty}^{\infty}\omega d\omega\text{Tr}[\mathbf{\Lambda}_\text{L}\mathbf{G}^\dagger(\omega)], \\[1ex]
\label{eq:right_flux}
	\nonumber	J_\text{R}(t) 
	 =&\int_{-\infty}^{\infty}\omega^2 \frac{k_\text{B}T^{(0)}_\text{R}}{\pi}d\omega\text{Tr}[\mathbf{G}^\dagger(\omega)\mathbf{\Lambda}_\text{R}^T\mathbf{G}(\omega)\mathbf{\Lambda}_\text{R}]+\int_{-\infty}^{\infty}\omega^2 \frac{k_\text{B}T^{(0)}_\text{L}}{\pi}d\omega\text{Tr}[\mathbf{G}^\dagger(\omega)\mathbf{\Lambda}_\text{R}^T\mathbf{G}(\omega)\mathbf{\Lambda}_\text{L}] \\
	\nonumber  &-\int_{-\infty}^{\infty}\omega (\omega+\omega_\text{R})e^{-i\omega_\text{R} t}\frac{k_\text{B}\Delta T_\text{R}}{2\pi i}d\omega\text{Tr}[\mathbf{G}^\dagger(\omega+\omega_\text{R})\mathbf{\Lambda}_\text{R}^T\mathbf{G}(\omega)\mathbf{\Lambda}_\text{R}]   \\
	\nonumber & -\int_{-\infty}^{\infty}\omega (\omega+\omega_\text{L})e^{-i\omega_\text{L} t}\frac{k_\text{B}\Delta T_\text{L}}{2\pi i}d\omega\text{Tr}[\mathbf{G}^\dagger(\omega+\omega_\text{L})\mathbf{\Lambda}_\text{R}^T\mathbf{G}(\omega)\mathbf{\Lambda}_\text{L}] \\
	\nonumber 	&+\int_{-\infty}^{\infty}\omega (\omega-\omega_\text{R})e^{i\omega_\text{R} t}\frac{k_\text{B}\Delta T_\text{R}}{2\pi i}d\omega\text{Tr}[\mathbf{G}^\dagger(\omega-\omega_\text{R})\mathbf{\Lambda}_\text{R}^T\mathbf{G}(\omega)\mathbf{\Lambda}_\text{R}] 	\\
	\nonumber  &+\int_{-\infty}^{\infty}\omega (\omega-\omega_\text{L})e^{i\omega_\text{L} t}\frac{k_\text{B}\Delta T_\text{L}}{2\pi i}d\omega\text{Tr}[\mathbf{G}^\dagger(\omega-\omega_\text{L})\mathbf{\Lambda}_\text{R}^T\mathbf{G}(\omega)\mathbf{\Lambda}_\text{L}]  \\ 
	&+i\frac{k_\text{B}T_\text{R}(t)}{\pi}\int_{-\infty}^{\infty}\omega d\omega\text{Tr}[\mathbf{\Lambda}_\text{R}\mathbf{G}^\dagger(\omega)], \\[1ex]
\label{eq:system_flux}
	\nonumber J_\text{sys}(t)
	=&-\int_{-\infty}^{\infty}\omega^2 \frac{k_\text{B}T^{(0)}_\text{L}}{\pi}d\omega\text{Tr}[\mathbf{G}^\dagger(\omega)\mathbf{\Lambda}^T\mathbf{G}(\omega)\mathbf{\Lambda}_\text{L}] -\int_{-\infty}^{\infty}\omega^2 \frac{k_\text{B}T^{(0)}_\text{R}}{\pi}d\omega\text{Tr}[\mathbf{G}^\dagger(\omega)\mathbf{\Lambda}^T\mathbf{G}(\omega)\mathbf{\Lambda}_\text{R}] \\ \nonumber
	&+\int_{-\infty}^{\infty}\omega (\omega+\omega_\text{L})e^{-i\omega_\text{L} t}\frac{k_\text{B}\Delta T_\text{L}}{2\pi i}d\omega\text{Tr}[\mathbf{G}^\dagger(\omega+\omega_\text{L})\mathbf{\Lambda}^T\mathbf{G}(\omega)\mathbf{\Lambda}_\text{L}]  \\ \nonumber
	&-\int_{-\infty}^{\infty}\omega (\omega-\omega_\text{L})e^{i\omega_\text{L} t}\frac{k_\text{B}\Delta T_\text{L}}{2\pi i}d\omega\text{Tr}[\mathbf{G}^\dagger(\omega-\omega_\text{L})\mathbf{\Lambda}^T\mathbf{G}(\omega)\mathbf{\Lambda}_\text{L}]  \\ \nonumber
	&+\int_{-\infty}^{\infty}\omega (\omega+\omega_\text{R})e^{-i\omega_\text{R} t}\frac{k_\text{B}\Delta T_\text{R}}{2\pi i}d\omega\text{Tr}[\mathbf{G}^\dagger(\omega+\omega_\text{R})\mathbf{\Lambda}^T\mathbf{G}(\omega)\mathbf{\Lambda}_\text{R}]  \\ \nonumber
	&-\int_{-\infty}^{\infty}\omega (\omega-\omega_\text{R})e^{i\omega_\text{R} t}\frac{k_\text{B}\Delta T_\text{R}}{2\pi i}d\omega\text{Tr}[\mathbf{G}^\dagger(\omega-\omega_\text{R})\mathbf{\Lambda}^T\mathbf{G}(\omega)\mathbf{\Lambda}_\text{R}]  \\ 
	&-i\frac{k_\text{B}}{\pi} \left(T_\text{L}(t)\int_{-\infty}^{\infty}\omega d\omega\text{Tr}[\mathbf{\Lambda}_\text{L}\mathbf{G}^\dagger(\omega)]
	+T_\text{R}(t)\int_{-\infty}^{\infty}\omega d\omega \text{Tr}[\mathbf{\Lambda}_\text{R}\mathbf{G}^\dagger(\omega)]\right),
\end{align}
\end{widetext}
where $\mathbf{\Lambda}=\mathbf{\Lambda}_\text{L}+\mathbf{\Lambda}_\text{R}$.
Here, we have used the conservation of energy relation $J_\text{L}(t)+J_\text{R}(t)+J_\text{sys}(t)=0$
to deduce the system energy flux.
Conservation of energy has been previously verified directly for the simpler case of a single-particle chain. \cite{craven2023b}
Equations~(\ref{eq:left_flux}), (\ref{eq:right_flux}), and (\ref{eq:system_flux}) are the primary analytical results of this article. They are the time-dependent energy flux expressions for the left bath, right bath, and the system in the presence of an oscillating temperature gradient.
In the limit of a single particle between two oscillating heat baths, the energy fluxes can be expressed in closed form  using the developed formalism (see Appendix~\ref{sec:oneparticleheatcurrent}).

\subsubsection{Quasistatic Limit}

The quasistatic (QS) limit is the limit in which the baths oscillate slowly with respect to the system-bath couplings  ($\omega_\text{L}/\gamma, \omega_\text{R}/\gamma \rightarrow 0,0)$.
In the QS limit, Eqs.~(\ref{eq:v_v_l}) and (\ref{eq:v_v_r}) become
\begin{align}
\label{eq:L}
 \nonumber	\left\langle(\mathbf{\Lambda}_\text{L}\dot{\mathbf{X}})^T \dot{\mathbf{X}} \right\rangle
	&=\frac{k_\text{B}T_\text{L}(t)}{\pi}\int_{-\infty}^{\infty}\omega^2 \text{Tr}[\mathbf{G}^\dagger(\omega)\mathbf{\Lambda}_\text{L}^T\mathbf{G}(\omega)\mathbf{\Lambda}_\text{L}] d\omega \\
	& +\frac{k_\text{B}T_\text{R}(t)}{\pi}\int_{-\infty}^{\infty}\omega^2 \text{Tr}[\mathbf{G}^\dagger(\omega)\mathbf{\Lambda}_\text{L}^T\mathbf{G}(\omega)\mathbf{\Lambda}_\text{R}] d\omega, \\
\label{eq:R}
 \nonumber	\left\langle(\mathbf{\Lambda}_\text{R}\dot{\mathbf{X}})^T \dot{\mathbf{X}} \right\rangle
	&=\frac{k_\text{B}T_\text{R}(t)}{\pi}\int_{-\infty}^{\infty}\omega^2 \text{Tr}[\mathbf{G}^\dagger(\omega)\mathbf{\Lambda}_\text{R}^T\mathbf{G}(\omega)\mathbf{\Lambda}_\text{R}] d\omega \\
	& +\frac{k_\text{B}T_\text{L}(t)}{\pi}\int_{-\infty}^{\infty}\omega^2 \text{Tr}[\mathbf{G}^\dagger(\omega)\mathbf{\Lambda}_\text{R}^T\mathbf{G}(\omega)\mathbf{\Lambda}_\text{L}] d\omega.
\end{align}
These expressions can be derived by noting that terms such as $(\omega \pm \omega_\text{L})$ and $(\omega \pm \omega_\text{R})$ in 
Eqs.~(\ref{eq:left_flux}) and (\ref{eq:right_flux}) reduce to simply $\omega$ in the QS limit, which can be shown by multiplying and dividing these terms by $\gamma$ and then evaluating those terms in the QS limit.

The QS limits of the noise-velocity correlation functions in Eqs.~(\ref{eq:v_xi_l}) and (\ref{eq:v_xi_r}) are
\begin{align}
	\nonumber &\left\langle\dot{\mathbf{X}}^T \mathbf{\Xi}_\text{L} \right\rangle
	=-i\frac{k_\text{B}T_\text{L}(t)}{\pi}\int_{-\infty}^{\infty}\omega \text{Tr}[\mathbf{\Lambda}_\text{L}\mathbf{G}^\dagger(\omega)]d\omega\\ 
	& \quad =\frac{k_\text{B}T_\text{L}(t)}{\pi}  \int_{-\infty}^{\infty}\omega^2 \text{Tr}[\mathbf{\Lambda}_\text{L}\mathbf{G}(\omega)(\mathbf{\Lambda}_\text{L}+\mathbf{\Lambda}_\text{R})\mathbf{G}^\dagger(\omega)]d\omega, \\
	\nonumber &\left\langle\dot{\mathbf{X}}^T \mathbf{\Xi}_\text{R} \right\rangle
	=-i\frac{k_\text{B}T_\text{R}(t)}{\pi}\int_{-\infty}^{\infty}\omega \text{Tr}[\mathbf{\Lambda}_\text{R}\mathbf{G}^\dagger(\omega)]d\omega\\ \nonumber
	& \quad =\frac{k_\text{B}T_\text{R}(t)}{\pi}  \int_{-\infty}^{\infty}\omega^2 \text{Tr}[\mathbf{\Lambda}_\text{R}\mathbf{G}(\omega)(\mathbf{\Lambda}_\text{R}+\mathbf{\Lambda}_\text{L})\mathbf{G}^\dagger(\omega)]d\omega, \\ 
\end{align}
where we have used the relation $\mathbf{G}^\dagger-\mathbf{G}=2i\mathbf{G}^\dagger\mathbf{\Gamma}\mathbf{G}$.

Applying the relation $\mathbf{\Lambda}_{\text{L}/\text{R}}^T=\mathbf{\Lambda}_{\text{L}/\text{R}}$ in Eqs.~(\ref{eq:L}) and (\ref{eq:R}), we now have expressions for the heat bath energy fluxes in the QS limit:
\begin{align}
	\label{eq:current_Left}
	\nonumber J^{\text{(QS)}}_\text{L}(t) &= \left\langle (\mathbf{\Lambda}_\text{L}\dot{\mathbf{X}})^T \dot{\mathbf{X}}\right\rangle  - \left\langle\dot{\mathbf{X}}^T\mathbf{\Xi}_\text{L}\right\rangle \\ 
	\nonumber &=\frac{k_\text{B}}{\pi}\big(T_\text{R}(t) -T_\text{L}(t)\big) \\ 
	&\quad \times \int_{-\infty}^{\infty}\omega^2 \text{Tr}[\mathbf{G}^\dagger(\omega)\mathbf{\Lambda}_\text{L}^T\mathbf{G}(\omega)\mathbf{\Lambda}_\text{R}]d\omega, \\[1ex]
	\label{eq:current_Right}
	\nonumber J^{\text{(QS)}}_\text{R}(t) &= \left\langle (\mathbf{\Lambda}_\text{R}\dot{\mathbf{X}})^T \dot{\mathbf{X}}\right\rangle - \left\langle\dot{\mathbf{X}}^T\mathbf{\Xi}_\text{R}\right\rangle \\ 
	\nonumber &=\frac{k_\text{B}}{\pi}\big(T_\text{L}(t) -T_\text{R}(t)\big) \\ 
	&\quad \times \int_{-\infty}^{\infty}\omega^2 \text{Tr}[\mathbf{G}^\dagger(\omega)\mathbf{\Lambda}_\text{R}^T\mathbf{G}(\omega)\mathbf{\Lambda}_\text{L}]d\omega.
\end{align}
Applying 
the identities $\mathbf{\Gamma}_\text{L}=\omega\mathbf{\Lambda}_\text{L}$ and $\mathbf{\Gamma}_\text{R}=\omega\mathbf{\Lambda}_\text{R}$,
we can write the left energy flux as
\begin{align}
\label{eq:heatcurrentquasiL}
\nonumber J^{\text{(QS)}}_\text{L}(t)  &= \frac{k_\text{B}\big(T_\text{R}(t) - T_\text{L}(t)\big)}{\pi} \\ 
&\quad \times \int_{-\infty}^\infty \text{Tr}[\mathbf{G}^\dagger(\omega)\mathbf{\Gamma}_\text{L}\mathbf{G}(\omega)\mathbf{\Gamma}_\text{R}]d\omega ,
\end{align}
and the right energy flux as
\begin{align}
\label{eq:heatcurrentquasiR}
\nonumber J^{\text{(QS)}}_\text{R}(t)  &= \frac{k_\text{B}\big(T_\text{L}(t) - T_\text{R}(t)\big)}{\pi}  \\ 
&\quad \times \int_{-\infty}^\infty \text{Tr}[\mathbf{G}^\dagger(\omega)\mathbf{\Gamma}_\text{L}\mathbf{G}(\omega)\mathbf{\Gamma}_\text{R}]d\omega.
\end{align}
The $\text{Tr}[\mathbf{G}^\dagger(\omega)\mathbf{\Gamma}_\text{L}\mathbf{G}(\omega)\mathbf{\Gamma}_\text{R}]$ integrand 
in $J^{\text{(QS)}}_\text{L}(t)$ and $J^{\text{(QS)}}_\text{R}(t)$
is the well-known expression for the static phononic transmission function.
Therefore, we see that in the QS limit the heat current through the system is simply the product of a factor containing the time-dependent temperature difference between baths and a frequency space integral over the same transmission function that would be obtained in the limit of a static temperature gradient.
Also note that in the QS limit, $J^{\text{(QS)}}_\text{L}(t) = -J^{\text{(QS)}}_\text{R}(t)$ for all $t$
because the system energy flux diminishes to zero, $J^{\text{(QS)}}_\text{sys} = 0$.
This is because the energy in the system is redistributed to the baths on a much faster timescale than the temperature oscillations.

\subsubsection{Static Limit}
In the limit of vanishing temperature oscillations ($\Delta T_\text{L}, \Delta T_\text{R} \rightarrow 0,0$), we arrive at the limit of a static (S) temperature gradient. In this limit, the temperature difference between baths is not oscillating.
The energy fluxes in the static limit derived using the developed formalism are
\begin{align}
\label{eq:heatcurrentquasi}
J^{\text{(S)}}_\text{L} &= -J^{\text{(S)}}_\text{R}\\
\nonumber &= \frac{k_\text{B}\left(T^{(0)}_\text{R} - T^{(0)}_\text{L}\right)}{\pi} \int_{-\infty}^\infty \text{Tr}[\mathbf{G}^\dagger(\omega)\mathbf{\Gamma}_\text{L}\mathbf{G}(\omega)\mathbf{\Gamma}_\text{R}]d\omega \\
 J^{\text{(S)}}_\text{sys} &= 0
\end{align}
which are well-known expressions that can be obtained using Landauer's formalism for heat conduction under steady-state conditions in the classical limit. \cite{Segal2016,Dhar2008,sharony2020stochastic}

\section{\label{sec:results}Results and Discussion}

\begin{figure}[t]
\includegraphics[width = 8.5cm,clip]{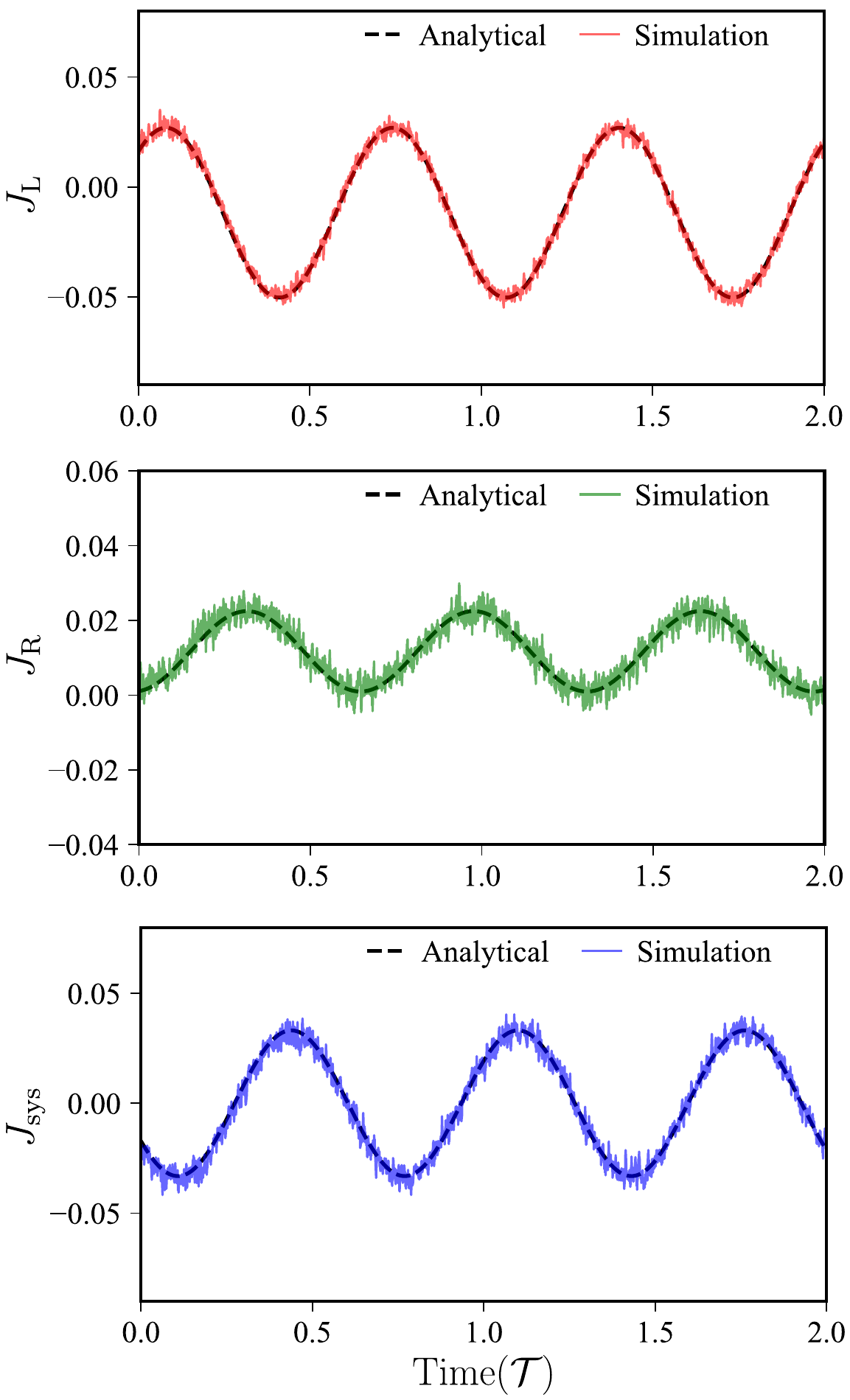}
\caption{\label{fig:Flux_1}
Time-dependence of the energy fluxes for the left bath (top), right bath (middle), and the system (bottom) 
in the case of a molecular chain consisting of $N=5$ particles connected to two heat baths with oscillating temperatures.
In each panel, the black dashed curve is the exact analytical result and the colored line is the result generated from molecular dynamics simulations.
Each energy flux is shown in units of $\gamma k_\text{B} T$.
Time is shown in units of the total oscillation period $\mathcal{T}$.
Parameters are $\gamma = 2$ ($\gamma_\text{L} = 1$, $\gamma_\text{R} = 1$), $m = 1$, $T^{(0)}_\text{L} = 1$, $T^{(0)}_\text{R} = 1.05$,
$\Delta T_\text{L} = 0.1$, $\Delta T_\text{R} = 0$, $\omega_\text{L} =5$, $\omega_\text{R} = 0$, $k = 250$, $k_\text{pin} = 0$.
All parameters throughout are given in reduced units with 
characteristic dimensions: $\widetilde{\sigma} = 1\,\text{\AA}$,  $\widetilde{\tau} = 1\,\text{ps}$,
$\widetilde{m} = 10\,m_u$,
and $\widetilde{T} = 300\,\text{K}$.
The $y$-axis values are running averages calculated from sets of 10 consecutive data points to reduce noise.
}
\end{figure}

To verify the accuracy of the derived theoretical expressions for the energy fluxes,
we performed molecular dynamics (MD) simulations of the harmonic chain model defined in Eq.~(\ref{eq:EoM1}) and measured the energy fluxes in those simulations. 
The specific method we used is the stochastic nonequilibrium dynamical simulation approach described in Ref.~\citenum{sharony2020stochastic}.
A comparison between the simulation results and the theoretical predictions is shown in Fig.~\ref{fig:Flux_1}.
The theoretical predictions are in excellent agreement for the left, right, and system energy fluxes under the 
chosen system and bath parameters values. 
We have verified that similar agreement is observed for other system parameters and chain lengths.
It is important to note that in the results shown in Fig.~\ref{fig:Flux_1}, the right bath temperature is not oscillating (i.e. $\Delta T_\text{R} = 0$)---only the left bath is oscillating---but the right heat bath still participates in the heat conduction process by giving and absorbing energy from the system.
This is shown in the middle panel of Fig.~\ref{fig:Flux_1} where the right bath energy flux is oscillatory
because of time-periodic energy propagation from the oscillating left bath. 

\begin{figure}[t]
\includegraphics[width = 8.5cm,clip]{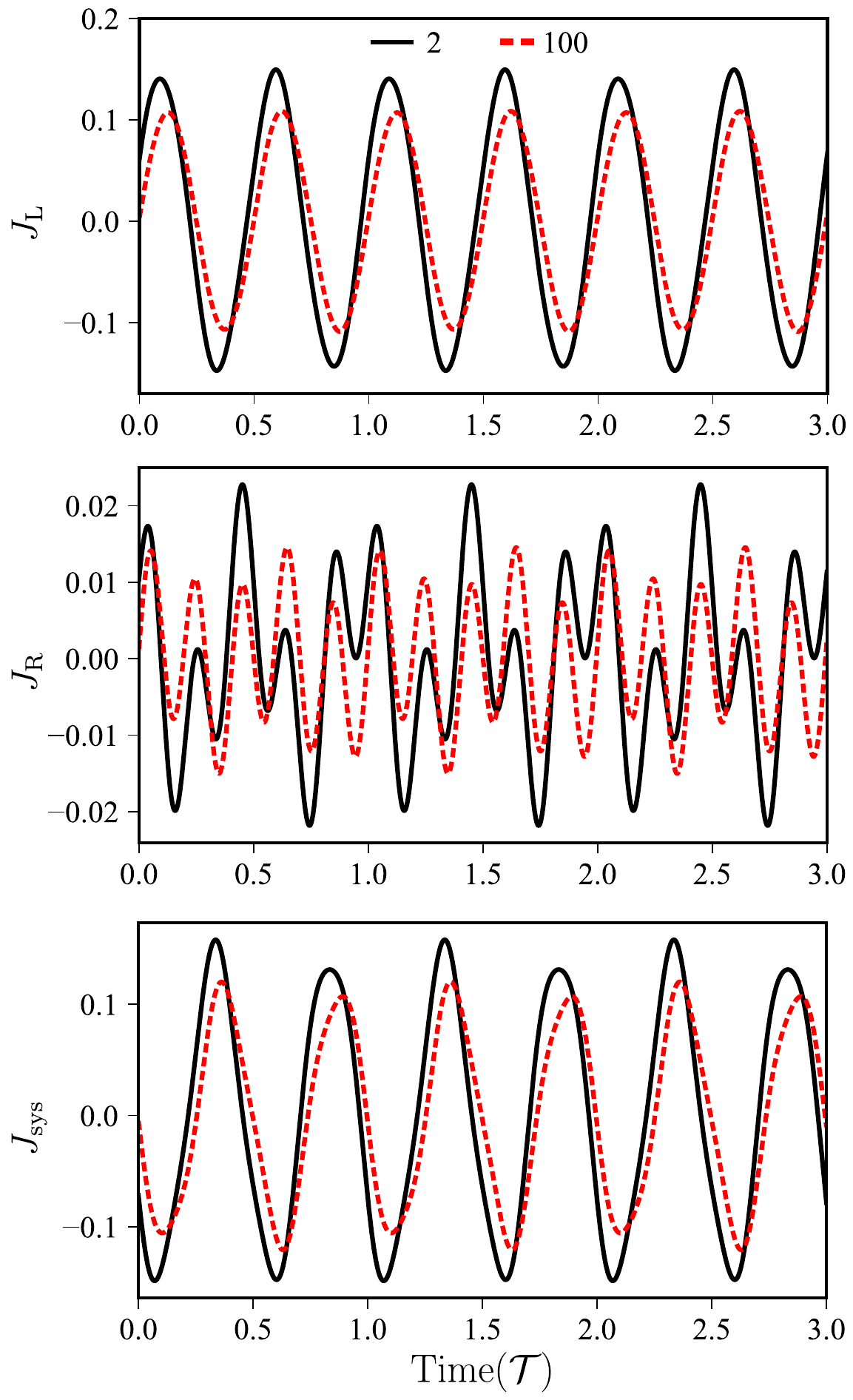}
\caption{\label{fig:len_currents}
Time-dependence of the energy fluxes for the left bath (top), right bath (middle), and the system (bottom) 
in a molecular chain consisting of $N=2$ (solid black) and $N = 100$ (dashed red) particles connected to two heat baths with oscillating temperatures.
Each energy flux is shown in units of $\gamma k_\text{B} T$.
Time is shown in units of the total oscillation period $\mathcal{T}$.
Parameters are $\gamma = 1.7$ ($\gamma_\text{L} = 1.5$, $\gamma_\text{R} = 0.2$), $m = 1$, $T^{(0)}_\text{L} = 1$, $T^{(0)}_\text{R} = 1$,
$\Delta T_\text{L} = 0.2$, $\Delta T_\text{R} = 0.1$, $\omega_\text{L} =2$, $\omega_\text{R} = 5$, $k = 250$, $k_\text{pin} = 0$.
All parameter values are given in reduced units as specified in the Fig.~\ref{fig:Flux_1} caption.}
\end{figure}

The time-dependent energy fluxes for two different chain lengths are shown in Fig.~\ref{fig:len_currents}.
The presented results provide insight into the combined effect of temperature oscillations and variation in the length of the molecular chain. 
Because of the oscillations, there is a time-periodic temperature gradient $\nabla T(t)  = \Delta T(t)/N$ across the system.
There are two key observations:
(a) complex patterns arise when the temperatures of both baths are oscillating in time due to the multiple frequencies and accompanying resonances that arise in the system and 
(b) small differences are observed in the energy fluxes when the chain length increases from $N=2$ (solid black) particles to $N=100$ (dashed red) particles, suggesting that ballistic transport dominates for purely harmonic chains even in the presence of a oscillatory temperature gradient.

\begin{figure}[t]
\includegraphics[width = 8.5cm,clip]{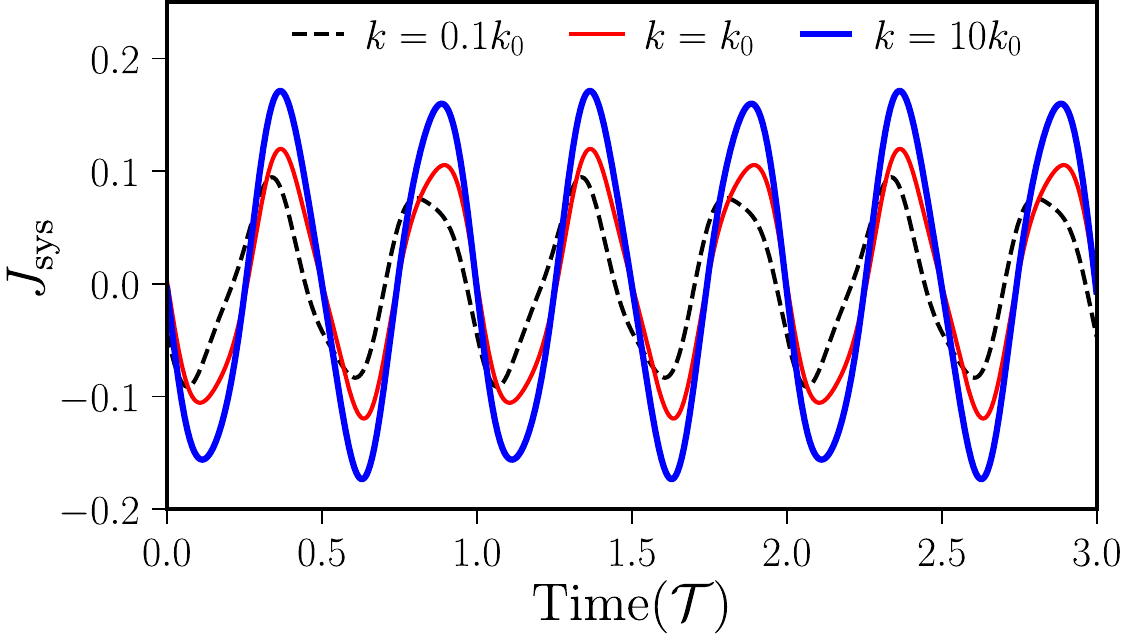}
\caption{\label{fig:sys_heat_kt}
Time-dependence of the energy flux for a system
consisting of $N=60$ particles connected by harmonic forces. The different curves correspond to the different force constants shown in the figure legend.
Each energy flux is shown in units of $\gamma k_\text{B} T$.
Time is shown in units of the total oscillation period $\mathcal{T}$.
Parameters are $\gamma = 1.7$ ($\gamma_\text{L} = 1.5$, $\gamma_\text{R} = 0.2$), $m = 1$, $T^{(0)}_\text{L} = 1$, $T^{(0)}_\text{R} = 1$,
$\Delta T_\text{L} = 0.2$, $\Delta T_\text{R} = 0.1$, $\omega_\text{L} =2$, $\omega_\text{R} = 5$, $k = k_0 = 250$, $k_\text{pin}= 0$.
All parameter values are given in reduced units as specified in the Fig.~\ref{fig:Flux_1} captions.
}
\end{figure}

A natural question that arises is 
how heat conduction is affected by the strength of the interatomic interactions (i.e., the force constants) 
within the molecular chain under external temperature driving. 
Figure~\ref{fig:sys_heat_kt} illustrates the system energy flux for different force constants in a chain of $N=60$ particles. 
The force constant in the model is denoted by $k$ and, for ease of exposition, we use a baseline force constant value of $k_0=250$ and express changes in the force constant as multiplicative factors of this value. 
The force constants in the model are always the same for each bond in the molecular chain. 
As shown in Fig.~\ref{fig:sys_heat_kt}, the stronger the interatomic interaction, 
the larger the maximum magnitude of the energy fluxes are. 
For example,  $k=10k_0$ exhibits approximately a 50\% increase in energy flux magnitude compared to $k=0.1k_0$. 
One might naively attribute this observation to stronger correlations between the atoms in the molecule when the the force constant is larger. 
However, as we will demonstrate later, 
this interpretation does not account for the complex interplay
between system-bath interactions.

\begin{figure}[t]
\includegraphics[width = 8.5cm,clip]{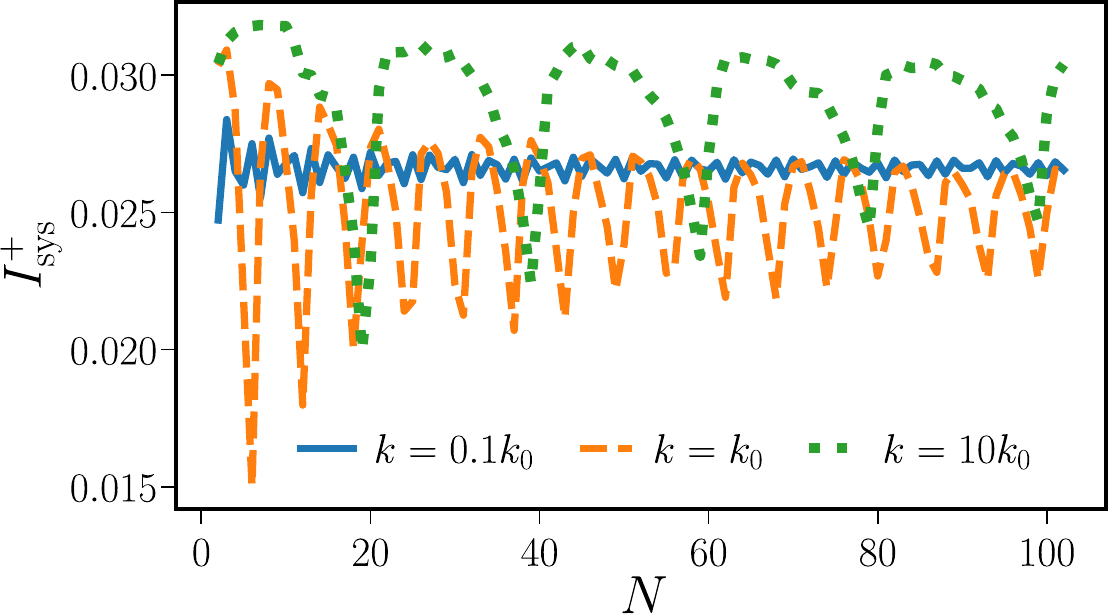}
\caption{\label{fig:sys_energy_N}
Average energy storage capacity calculated over one period of oscillation as a function of chain length for the three different force constant values shown in the legend.
The capacity is calculated using Eq.~\ref{eq:heatplus}. 
Time is shown in units of the total oscillation period $\mathcal{T}$.
The capacity is shown in units of $\gamma k_\text{B} T$.
Parameters are $\gamma = 1$ ($\gamma_\text{L} = 1$, $\gamma_\text{R} = 1$), $m = 1$, $T^{(0)}_\text{L} = 1$, $T^{(0)}_\text{R} = 1.5$,
$\Delta T_\text{L} = 0.1$, $\Delta T_\text{R} = 0$, $\omega_\text{L} =5$, $\omega_\text{R} = 0$, $k = k_0 = 250$, $k_\text{pin}= 0$.
All parameter values are given in reduced units as specified in the Fig.~\ref{fig:Flux_1} caption.
}
\end{figure}

Figure~\ref{fig:sys_energy_N} illustrates the length-dependent average energy storage capacity
of the system over one period of oscillation for varying force constants. 
The average energy storage capacity $I_\text{sys}$ is defined in Eq.~\ref{eq:heatplus}.
The fluctuating behavior is observed for three force constants and chain lengths, 
reflecting the oscillatory temperatures from the baths. 
However, these fluctuations are not strictly in the time domain and can be nonlinear and oscillatory when other system variables are varied besides time.
Several observations are of note: 
(a) It is not a straightforward assertion
that stronger interatomic interactions result in higher conduction.
For instance, specific chain lengths exhibit smaller energy storage capacities for $k=10k_0$ 
compared to $k=0.1k_0$, 
although, in general, the values for $k=10k_0$ are higher. 
(b) Smaller force constants appear to yield smaller oscillating magnitudes. 
This trend can be attributed to larger force constants increasing the upper limit 
of frequencies of hybridized modes within the chains which
amplifies conduction through different channels in comparison to cases with smaller force constants. 
(c) There are discernible trends for all three heat current intensities to approach certain plateaus. 
They commence with larger fluctuations and stabilize as the chain lengths increase. 
This observation aligns with steady-state 
length-dependent heat conduction principles, 
where the thermal conductance of long chains becomes length-independent. \cite{segal2003thermal}  
A significant feature in this work is 
that the oscillations of the energy storage capacities do not vanish, 
but instead fluctuate about certain baseline values.

\begin{figure}[t]
\includegraphics[width = 8.5cm,clip]{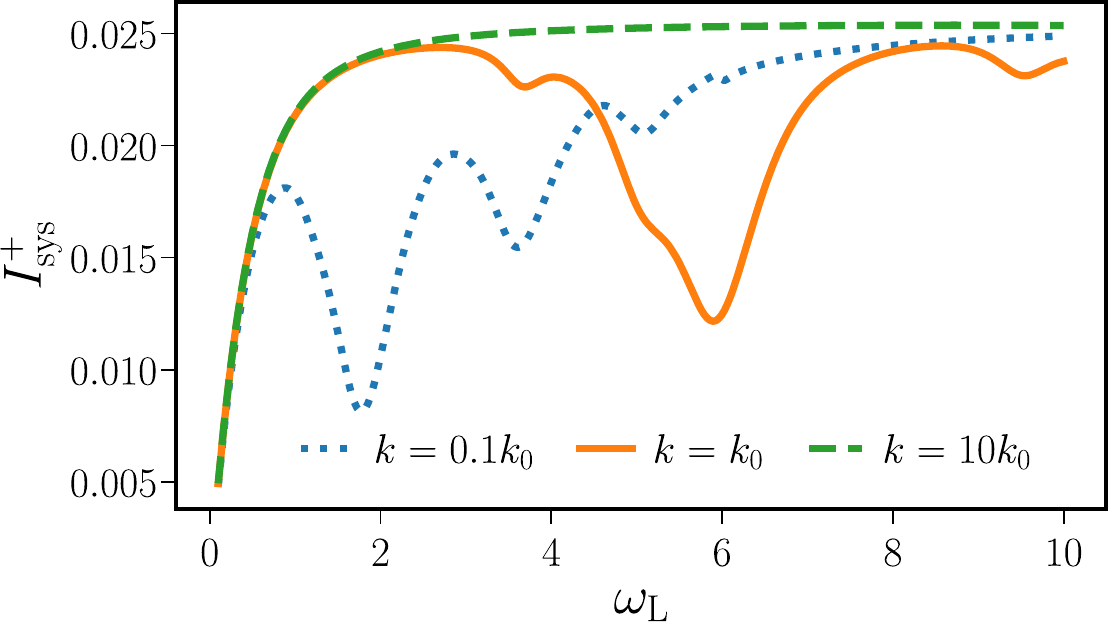}
\caption{\label{fig:sys_energy_omegaL}
Average energy storage capacity calculated over one period of oscillation as a function of left temperature oscillating frequency $\omega_\text{L}$ for chain length of $N=5$, under three different force constants ($k$).
The capacity is calculated using Eq.~\ref{eq:heatplus}. 
The capacity is shown in units of $\gamma k_\text{B} T$.
Time is shown in units of the total oscillation period $\mathcal{T}$.
Parameters are $\gamma = 2$ ($\gamma_\text{L} = 1$, $\gamma_\text{R} = 1$), $m = 1$, $T^{(0)}_\text{L} = 1$, $T^{(0)}_\text{R} = 1.5$,
$\Delta T_\text{L} = 0.1$, $\Delta T_\text{R} = 0$, $\omega_\text{L} =5$, $\omega_\text{R} = 0$, $k_0 = 250$, $k_\text{pin}= 0$.
All parameter values are given in reduced units as specified in the Fig.~\ref{fig:Flux_1} caption.
}
\end{figure}

Interesting transport phenomena emerge as the driving frequencies of the baths are modified, 
as demonstrated in Fig.~\ref{fig:sys_energy_omegaL}
which illustrates the system's average energy storage capacity over different oscillation frequencies. 
For simplicity, we consider the case in which only the left temperature
is oscillating (i.e., $\Delta T_\text{R}=0$).
The dips in the curves signify resonant states between vibrational frequencies in the system and bath oscillation frequencies. 
Taking the case of $k = k_0$ (corresponding to $k_0/m \sim 5 \text{ps}^{-1}$ in our example), 
the first significant dip occurs around $\omega_\text{L} \sim  5 \text{ps}^{-1}$, 
suggesting a hybridized mode around the $k_0$ force constant frequency could be contributing the transport. 
The trends for smaller $k$ values exhibit earlier occurrences of dips, 
while when $k_0$ is increased by a factor of ten above the baseline value, resonances under $\omega_\text{L}=10$ values are not observed. We have confirmed that dips do occur when $\omega_\text{L}>10$.
One observation regarding higher driving frequencies is that when the bath oscillates extremely fast, 
the average energy storage capacities for all three force constants examined appear to converge to the same asymptotic value.   
This implies that the energy storage capacities for the system gradually become agnostic to the force constant value.
Conversely, at the opposite end of the spectrum, 
when the bath temperatures oscillate extremely slowly, 
the energy storage capacity diminishes to zero. 
This is a result of the vanishing system energy fluxes 
as the system gradually approaches the quasistatic limit. 
The example in Fig.~\ref{fig:sys_energy_omegaL} is presented for $N=5$ particles, 
but similar behaviors are observed for longer chains.

\begin{figure}[t]
\includegraphics[width = 8.5cm,clip]{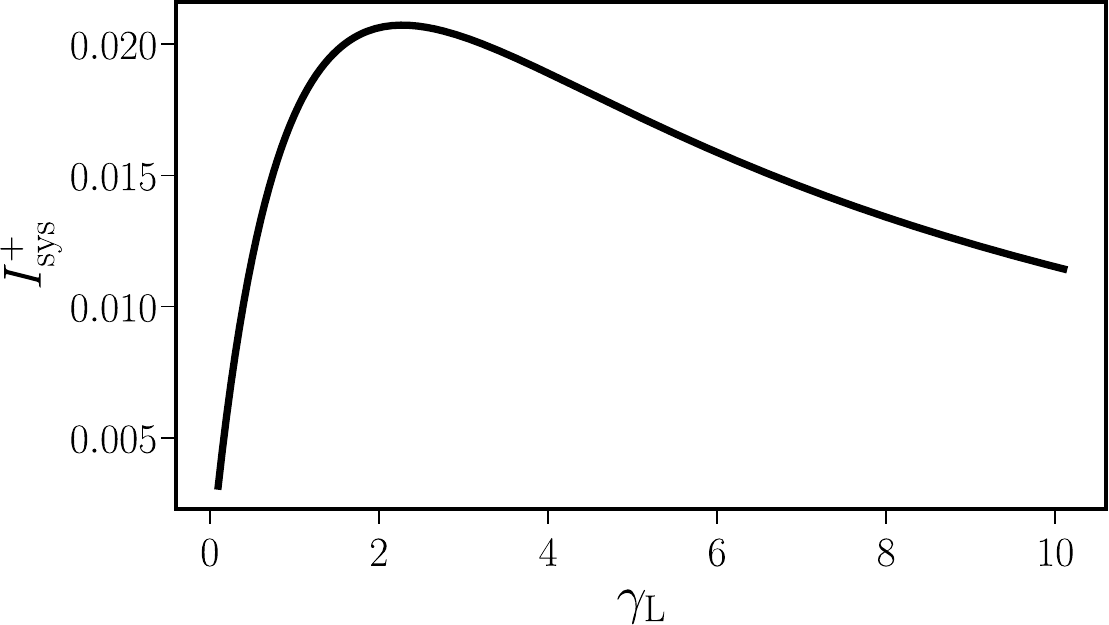}
\caption{\label{fig:sys_energy_gamma}
Average energy storage capacity as function of system-bath coupling for a chain length of $N=5$ and a force constant $k= k_0 = 250$.
The capacity is calculated using Eq.~\ref{eq:heatplus}
and is shown in units of $\gamma k_\text{B}T$, a unit which is changing as $\gamma_\text{L}$ is varied.
Time is shown in units of the total oscillation period $\mathcal{T}$.
Parameters are $m = 1$, $T^{(0)}_\text{L} = 1$, $T^{(0)}_\text{R} = 1.5$, $\gamma_\text{R} = 1$,
$\Delta T_\text{L} = 0.1$, $\Delta T_\text{R} = 0$, $\omega_\text{L} =5$, $\omega_\text{R} = 0$, $k_\text{pin}= 0$.
All parameter values are given in reduced units as specified in the Fig.~\ref{fig:Flux_1} caption.
}
\end{figure}

Figure~\ref{fig:sys_energy_gamma} illustrates 
the storage capacity as a function of the left system-bath coupling ($\gamma_\text{L}$). 
The right system-bath coupling is held constant at $\gamma_\text{R}=1$
and the force constants inside the chain (composed of $N=5$ particles) 
are uniformly set to $k = k_0$. 
When the coupling on the left is extremely small, 
indicating that the system is nearly decoupled from the left bath, 
the energy storage vanishes. 
Conversely, for strong system-bath coupling, 
the capacity for the system itself to retain energy diminishes gradually because heat conduction through the system occurs quickly due to the strong coupling. The implication of this is that heat moves more quickly through the system and less energy is retained. 
This nonlinear pattern once again highlights the complicated interplay 
between all factors (dynamical and system-specific) in the transient thermal transport mechanism examined in this work, as elucidated by our analytical results.

\section{\label{sec:conclusions}Conclusions}
We have examined vibrational heat transport through a harmonic molecular chain in the presence of a 
temperature gradient that is oscillating in time.
Using a nonequilibrium Green's function approach modified to account for the time-dependent thermal gradient, we have developed a theoretical framework to describe the heat transport properties in the examined system analytically.
The presented results demonstrate that the time-periodic modulation of a temperature gradient can significantly affect the heat transport and energy storage properties of nanoscale systems.

We have specifically shown that the introduction of an oscillating temperature gradient modifies the thermal conductance through a molecular lattice leading to altered heat transport, energy storage, and energy release.
These modifications arise due to differences in magnitude between the system-bath relaxation rates and other frequencies in the system, for example the oscillation frequency of the temperature gradient.
By tuning the properties of the oscillating gradient, such as its waveform, frequency, and amplitude,  
it is possible to manipulate the amount of energy stored within the system.
This effect has potential applications for improved energy harvesting and energy storage technologies, such as thermal batteries.

While the work in this article focuses on establishing a mathematical framework to describe heat transport under a time-periodic temperature gradient, the effects we have predicted could potentially be observed experimentally using current set-ups capable of measuring heat transport at the single-molecule level. 
Specific systems that may exhibit the phenomena we have predicted include alkanedithiol chains and polymeric molecular systems with metallic substrates in molecular junction set ups under periodic temperature oscillations. \cite{cui2019thermal,dinpajooh2020heat,dinpajooh2022heat}

Investigating how anharmonicities in the lattice structure affect heat transport is a potentially important next step.
Our preliminary findings suggest that when the system is asymmetric and the interaction forces between particles are anharmonic, the time-periodic modulation of a temperature gradient can give rise to emergent rectification effects. Examining these effects is a focus of our current work.
Applying nonequilibrium thermodynamic relations \cite{Horowitz2020thermodynamic} and formalisms to separate the time-dependent currents into positive and negative components \cite{craven18a1,craven18a2,BaratoPRL2018} will provide further insight into the thermodynamic properties of the system.
The present work focuses on the classical limit of temperature-modulated heat conduction. An important future direction involves extending the model to incorporate quantum mechanical effects. 
This extension will be needed in order to understand the effects described in this article in regimes in which quantum effects alter transport, for example, at low temperatures.
Overall, the findings in this article illustrate that nanoscale systems can be tailored to exhibit controllable heat transport properties and controlled energy storage through the temporal modulation of a temperature gradient.

\section{Acknowledgements}
We acknowledge support from the Los Alamos National Laboratory (LANL) 
Directed Research and Development funds (LDRD).
This research was performed in part at the Center for Nonlinear Studies (CNLS) at LANL. 
The computing resources used to perform this
research were provided by the LANL Institutional Computing Program.


\appendix

\section{\label{sec:oneparticleheatcurrent} Heat current derivation for a single-particle ($N=1$) chain}
The developed theoretical framework can be applied to generate closed-form expressions for the energy fluxes in the 
case of a single particle bridging two heat baths with oscillating temperatures.
These results can then be compared and validated using results obtained previously through an approach that uses direct calculation of the time correlation functions in the energy flux expressions. \cite{craven2023b}

The equation of motion for a  single free particle is 
\begin{equation}
     m\ddot{x} = -m\gamma_\text{L}\dot{x}- m \gamma_\text{R}\dot{x} + \xi_\text{L}(t)  + \xi_\text{R}(t),
\end{equation}
which can be written as
\begin{equation}
    -m\omega^2\tilde{x} = -im\omega\gamma\tilde{x} + \tilde{\xi}(\omega),
\end{equation}
in Fourier space
where we have combined the effects of left and right baths into effective friction
$\gamma=\gamma_\text{L}+\gamma_\text{R}$ and effective noise $\tilde{\xi}(\omega)=\tilde{\xi}_\text{L}(\omega)+\tilde{\xi}_\text{R}(\omega)$
terms. 
Note that the noise-noise correlation functions are the same as in Eqs.~(\ref{eq:fluctuation_correlationL}) and (s\ref{eq:fluctuation_correlationR}).
The equation of motion can be written as 
\begin{equation}
\label{eq:Eomsingle}
    m\tilde{x}(\omega)=G(\omega)\tilde{\xi}(\omega),
\end{equation}
where
\begin{equation}
\label{eq:Green_oneparticle}
    G(\omega)=\frac{1}{-\omega^2+i\omega\gamma}.  
\end{equation}
For algebraic simplicity, we have kept the mass term on the LHS of Eq.~(\ref{eq:Eomsingle}) and therefore it does not appear in $G$ in the following derivation (compare with Eq.~(\ref{eq:Langevin_chain_FT_M}) where the mass is moved to the RHS).

The general energy flux expressions for the two heat baths are
\begin{align}
    J_\text{L}&= m \gamma_\text{L} \big\langle \dot{x}^2(t)\big\rangle-\big\langle \xi_\text{L}(t)\dot{x}(t) \big\rangle ,\\
    J_\text{R}&= m \gamma_\text{R} \big\langle \dot{x}^2(t)\rangle-\big\langle \xi_\text{R}(t)\dot{x}(t) \big\rangle.
\end{align}
The second moment of the particle velocity can be expressed as
\begin{widetext}
\begin{align}
\label{eq:v2appendix}
    \nonumber \big\langle \dot{x}^2(t)\big\rangle &= \frac{k_\text{B}\left(T^{(0)}_\text{L}\gamma_\text{L} +T^{(0)}_\text{R} \gamma_\text{R}\right)}{m\pi}\int_{-\infty}^{\infty} \omega^2|G(\omega)|^2d\omega \\ \nonumber
    &+i\frac{k_\text{B}\gamma_\text{L}\Delta T_\text{L}}{2m\pi}\int_{-\infty}^{\infty} \left[(\omega_\text{L}-\omega)G(\omega_\text{L}-\omega)e^{i\omega_\text{L}t}-(\omega+\omega_\text{L})G(-\omega_\text{L}-\omega)e^{-i\omega_\text{L}t}\right]\omega G(\omega)d\omega \\ 
    &+i\frac{k_\text{B}\gamma_\text{R}\Delta T_\text{R}}{2m\pi}\int_{-\infty}^{\infty} \left[(\omega_\text{R}-\omega)G(\omega_\text{R}-\omega)e^{i\omega_\text{R}t}-(\omega+\omega_\text{R})G^\dagger(-\omega_\text{R}-\omega)e^{-i\omega_\text{R}t}\right]\omega G(\omega)d\omega,
\end{align}
\end{widetext}
and the noise-velocity correlation functions can be expressed as
\begin{align}
\label{eq:xivappendixL}
    \nonumber \big\langle \xi_\text{L}(t)\dot{x}(t) \big\rangle&=i\frac{\gamma_\text{L} k_\text{B}T^{(0)}_\text{L}}{\pi}\int_{-\infty}^{\infty} \omega G(\omega) d\omega\\
    &+i\frac{\gamma_\text{L} k_\text{B}\Delta T_\text{L}\sin(\omega_\text{L} t)}{\pi}
    \int_{-\infty}^{\infty} \omega G(\omega)d\omega, \\
\label{eq:xivappendixR}
    \nonumber \big\langle \xi_\text{R}(t)\dot{x}(t) \big\rangle&=i\frac{\gamma_\text{R} k_\text{B}T^{(0)}_\text{L}}{\pi}\int_{-\infty}^{\infty} \omega G(\omega) d\omega\\
    &+i\frac{\gamma_\text{R} k_\text{B}\Delta T_\text{R}\sin(\omega_\text{R} t)}{\pi}
    \int_{-\infty}^{\infty} \omega G(\omega)d\omega .
\end{align}
The integrals in these expressions can be evaluated using the relations:
\begin{widetext}
\begin{align}
\label{eq:diff_integral}
     &\int_{-\infty}^{\infty} \omega^2| G(\omega)|^2 d\omega
    =\frac{\pi}{\gamma}\\ 
     &\int_{-\infty}^{\infty}  (\omega_\text{L}-\omega)\omega G(\omega) G(\omega_\text{L}-\omega) d\omega = -\frac{2\pi(2\gamma-i\omega_\text{L})}{4\gamma^2+\omega_\text{L}^2} \\ 
     &\int_{-\infty}^{\infty}  (\omega_\text{R}-\omega)\omega G(\omega) G(\omega_\text{R}-\omega) d\omega = -\frac{2\pi(2\gamma-i\omega_\text{R})}{4\gamma^2+\omega_\text{R}^2} \\ 
     &\int_{-\infty}^{\infty} (-\omega_\text{L}-\omega)\omega G(\omega) G(-\omega_\text{L}-\omega) d\omega = -\frac{2\pi(2\gamma+i\omega_\text{L})}{4\gamma^2+\omega_\text{L}^2}\\ 
     &\int_{-\infty}^{\infty} (-\omega_\text{R}-\omega)\omega G(\omega) G(-\omega_\text{R}-\omega) d\omega = -\frac{2\pi(2\gamma+i\omega_\text{R})}{4\gamma^2+\omega_\text{R}^2}\\ 
     &\int_{-\infty}^{\infty} \omega G(\omega) d\omega
    =-i\gamma\int_{-\infty}^{\infty} \omega^2| G(\omega)|^2d\omega=-i\pi.\label{eq:Green_int_middle}
\end{align}
\end{widetext}
Appendix~\ref{sec:oneparticleGreenIntegral} contains a detailed explanation for the middle step of Eq.~(\ref{eq:Green_int_middle}).
After some algebraic manipulation, Eqs.~(\ref{eq:diff_integral})-(\ref{eq:Green_int_middle}) can be plugged into Eqs.~(\ref{eq:v2appendix}), (\ref{eq:xivappendixL}), and (\ref{eq:xivappendixR}) to obtain
\begin{widetext}
\begin{align}
    \big\langle \dot{x}^2(t)\big\rangle =\frac{k_B}{m}\left( T + \frac{2\gamma_\text{L}\Delta T_\text{L}[2\gamma\sin(\omega_\text{L}t)-\omega_\text{L}\cos(\omega_\text{L}t)]}{4\gamma^2+\omega_\text{L}^2}
    +\frac{2\gamma_\text{R}\Delta T_\text{R}[2\gamma\sin(\omega_\text{R}t)-\omega_\text{R}\cos(\omega_\text{R}t)]}{4\gamma^2+\omega_\text{R}^2}\right),
\end{align}
\end{widetext}
and
\begin{align}
    \langle \xi_\text{L}(t)\dot{x}(t) \rangle &=\gamma_\text{L} k_\text{B} T_\text{L}(t),\\
    \langle \xi_\text{R}(t)\dot{x}(t) \rangle &=\gamma_\text{R} k_\text{B} T_\text{R}(t).
\end{align}
We have defined the effective temperature of the system as
\begin{equation}
T=\frac{\gamma_\text{L}T^{(0)}_\text{L}+\gamma_\text{R}T^{(0)}_\text{R}}{\gamma_\text{L}+\gamma_\text{R}}.
\end{equation}
Therefore we now have closed form expressions for the energy fluxes:
\begin{widetext}
\begin{align}
    J_\text{L}(t) &= 
\gamma_\text{L} k_\text{B} \bigg( T - T_\text{L} (t) 
+ \frac{2\gamma_\text{L} \Delta T_\text{L} (2 \gamma \sin(\omega_\text{L} t) - \omega_\text{L} \cos(\omega_\text{L} t))}{4 \gamma^2+\omega^2_\text{L}} 
+ \frac{2\gamma_\text{R} \Delta T_\text{R} (2 \gamma \sin(\omega_\text{R} t) - \omega_\text{R} \cos(\omega_\text{R} t))}{4 \gamma^2+\omega^2_\text{R} }
 \bigg), \\
 J_\text{R}(t) &=  
\gamma_\text{R} k_\text{B} \bigg( T -T_\text{R} (t) + \frac{2\gamma_\text{L} \Delta T_\text{L} (2 \gamma \sin(\omega_\text{L} t) - \omega_\text{L} \cos(\omega_\text{L} t))}{4 \gamma^2+\omega^2_\text{L}} 
+ \frac{2 \gamma_\text{R} \Delta T_\text{R} (2 \gamma \sin(\omega_\text{R} t) - \omega_\text{R} \cos(\omega_\text{R} t))}{4 \gamma^2+\omega^2_\text{R} }
 \bigg), \\
 J_\text{sys} &= k_\text{B}\bigg( \frac{  \gamma_\text{L} \Delta T_\text{L} \omega_\text{L} (2 \gamma \cos(\omega_\text{L} t) + \omega_\text{L} \sin(\omega_\text{L} t))}{4 \gamma^2+\omega^2_\text{L}} 
+ 
\frac{  \gamma_\text{R} \Delta T_\text{R} \omega_\text{R}  (2 \gamma \cos(\omega_\text{R} t) + \omega_\text{R} \sin(\omega_\text{R} t))}{4 \gamma^2+\omega^2_\text{R}}
 \bigg),
\end{align}
\end{widetext}
where we have used the conservation of energy relation $J_\text{L}(t)+J_\text{R}(t)+J_\text{sys}(t)=0$ to obtain $J_\text{sys}$.
The energy flux expressions derived here using the NEGF approach are in agreement with the results we have derived previously by directly evaluating the correlation functions for a single Brownian particle in Ref.~\citenum{craven2023b}.

\section{\label{sec:oneparticleGreenIntegral} Derivation for the relation used in Eq.~(\ref{eq:Green_int_middle})}
We want to derive the relation
\begin{equation}
    \int_{-\infty}^{\infty} \omega G(\omega) d\omega
    =-i\gamma\int_{-\infty}^{\infty} \omega^2| G(\omega)|^2d\omega.
\end{equation}
The complex conjugate of the Green's function 
\begin{equation}
    G(\omega)=\frac{1}{-\omega^2+i\omega\gamma},  
\end{equation}
in Eq.~(\ref{eq:Green_oneparticle}) is
\begin{equation}
\label{eq:Green_oneparticle_cojugate}
    G^-(\omega)=\frac{1}{-\omega^2-i\omega\gamma}.    
\end{equation}
Using the definition of $G(\omega)$, we have
\begin{align}
    \int_{-\infty}^{\infty} \omega G(\omega) d\omega 
    &=\int_{-\infty}^{\infty}\frac{1}{-\omega+i\gamma} d\omega \nonumber \\
     &=\int_{-\infty}^{\infty}\frac{-\omega-i\gamma}{\omega^2+\gamma^2} d\omega \nonumber \\
     &=-\int_{-\infty}^{\infty}\frac{i\gamma}{\omega^2+\gamma^2} d\omega, 
\end{align}
where we have used the fact that integration of an odd function is zero in the last step.
Similarly, we have
\begin{align}
    \int_{-\infty}^{\infty} \omega G^-(\omega) d\omega 
    &=\int_{-\infty}^{\infty}\frac{1}{-\omega-i\gamma} d\omega \nonumber \\
     &=\int_{-\infty}^{\infty}\frac{-\omega+i\gamma}{\omega^2+\gamma^2} d\omega \nonumber \\
     &=\int_{-\infty}^{\infty}\frac{i\gamma}{\omega^2+\gamma^2} d\omega. 
\end{align}
and we therefore now have the relation
\begin{equation}
    \int_{-\infty}^{\infty} \omega G(\omega) d\omega=-\int_{-\infty}^{\infty} \omega G^-(\omega) d\omega,
\end{equation}
which infers 
\begin{equation}
2\int_{-\infty}^{\infty} \omega G(\omega) d\omega=\int_{-\infty}^{\infty} \omega [G(\omega)-G^-(\omega)] d\omega.
\end{equation}
Using the identity $G-G^-=-2i\omega\gamma| G(\omega)|^2$,
we have arrived at Eq.~(\ref{eq:Green_int_middle})
\begin{equation}
    \int_{-\infty}^{\infty} \omega G(\omega) d\omega
    =-i\gamma\int_{-\infty}^{\infty} \omega^2| G(\omega)|^2d\omega.
\end{equation}

\bibliographystyle{apsrev}
\bibliography{main.bbl}

\end{document}